\documentclass[aps,preprintnumbers,nofootinbib,superscriptaddress]{revtex4}

\usepackage{graphicx}
\DeclareGraphicsExtensions{.pdf}

\usepackage{amsfonts,amssymb,amsmath}
\usepackage{amsthm}

\newcommand{\comment}[1]{}
\newcommand{\ket}[1]{\left |  #1 \right\rangle}
\newcommand{\bra}[1]{\left \langle #1  \right |}

\newcommand{\Tr}{{\rm Tr}}
\newcommand{\St}{{\cal S}}
\theoremstyle{plain}
\newtheorem{theorem}{Theorem}
\newtheorem{lemma}{Lemma}

\theoremstyle{definition}

\begin{document}

\title{Security Details for Bit Commitment by Transmitting
  Measurement Outcomes}
\author{Sarah \surname{Croke}}
\affiliation{Perimeter Institute for Theoretical Physics, 31 Caroline Street North, Waterloo, ON N2L 2Y5, Canada.}
\author{Adrian \surname{Kent}}
\affiliation{Centre for Quantum Information and Foundations, DAMTP, Centre for
  Mathematical Sciences, University of Cambridge, Wilberforce Road,
  Cambridge, CB3 0WA, U.K.}
\affiliation{Perimeter Institute for Theoretical Physics, 31 Caroline Street North, Waterloo, ON N2L 2Y5, Canada.}

\date{August 2012}

\begin{abstract}
We spell out details of a simple argument for a security bound for
the secure relativistic quantum bit commitment protocol of Ref. \cite{bcmeasurement}. 
\end{abstract}

\maketitle

{ \bf Introduction} 

Recently, a new quantum relativistic bit commitment
protocol \cite{bcmeasurement} was introduced.
Its security relies, 
essentially, on the impossibility of completing a nonlocal measurement on
a distributed state outside the joint future light cone of
its components.    
Its implementation requires minimal quantum resources: 
the receiver needs to send quantum states (which can be 
unentangled qubits) to the committer, who needs to carry out
individual measurements on them as soon as they are received. 
No further quantum commmunication is required by either party;
nor do they require any entanglement, collective measurements, or  
quantum state storage. 

We present the protocol here in an 
idealized form assuming perfect state preparations, transmissions
and measurements.  We also make idealizations about the relativistic geometry and 
signalling speed, supposing that Alice and Bob each have agents in secure
laboratories infinitesimally separated from the points $P$, $Q_0$ and
$Q_1$, Alice can signal at precisely light speed, and all information
processing is instantaneous.  
We discuss here the simplest version of the scheme using qubit states 
and measurements in the standard BB84 basis \cite{BBeightyfour}.

Alice and Bob agree on a space-time point $P$, a set of coordinates
$(x,y,z,t)$ for Minkowski 
space, with $P$ as the origin, and (in the simplest case)
two points $Q_0 = (x,0,0,x)$  and $Q_1 = (-x,0,0,x)$ light-like  
separated from $P$. 
They each have agents, separated in secure laboratories, adjacent to 
each of the points $P$, $Q_0$, $Q_1$.  To simplify for the moment,
we take the distances from the labs to the relevant points as
negligible.    

Bob securely prepares a set of qubits $\ket{\psi_i}_{i=1}^N$ independently
randomly chosen from the BB84 states $\{ \ket{0} , \ket{1} , \ket{+},
\ket{-} \}$ (where $\ket{\pm} = \frac{1}{\sqrt{2}} ( \ket{0} \pm
\ket{1} )$) and sends them to Alice to arrive (essentially) at $P$. 
To commit to the bit value $0$, Alice measures each state in the $\{ \ket{0} ,
\ket{1} \}$ basis, and sends the outcomes over secure classical 
channels to her agents at $Q_0$ and $Q_1$. 
To commit to $1$, Alice measures each state in the $\{ \ket{+} ,
\ket{-} \}$ basis, and sends the outcomes as above. 
Alice's secure classical channels could, for example, be created
by pre-sharing one-time pads between her agent at $P$ and those
at $Q_0$ and $Q_1$ and sending pad-encrypted classical signals.
If necessary or desired, these pads could be periodically replenished by 
quantum key distribution links between the relevant agents. 

To unveil her committed bit, Alice's agents at $Q_0$ and $Q_1$
reveal the measurement outcomes to Bob's agents there. 
After comparing the revealed data to check that the declared
outcomes on both wings are the same (somewhere in the intersection
of the future light cones of $Q_0$ and $Q_1$), and that both are 
consistent with the list of  states sent at $P$, Bob accepts the commitment
and unveiling as genuine.  If the declared outcomes are different,
Bob has detected Alice cheating. 

{\bf Security} \qquad The 
protocol is evidently secure against Bob, who learns nothing
about Alice's actions until (if) she chooses to unveil the bit. 

Alice is constrained in that she has to be able to
reveal her commitment data at both $Q_0$ and
$Q_1$, since Bob's agents at these points verify the 
timing and location of the unveilings, and then later
compare the data to check they are consistent. 
We need to show that, if she is able to do so then,
essentially (up to some small probability defined 
in terms of a security parameter) she
was committed at $P$.  (See Ref. \cite{qtasks} for a
more formal discussion of security in terms of a space-time
oracle model.)
 
By Minkowski causality, Alice's ability to
unveil data consistent with a $0$ or $1$ commitment at $Q_0$ depends
only on operations she carries out on the line $P Q_0$.  Suppose that
she has a strategy in which she carries out some operations at $P$,
but these leave her significantly uncommitted, in the sense that her
optimal strategies ${\St}_i$ for successfully unveiling the bit values
$i$, by carrying out suitable operations in the causal future of $P$,
have success probabilities $p_i$, with $p_0 + p_1 > 1 + \delta$, for
some $\delta >0$.  By Minkowksi causality, any operations she carries
out on the half-open line segment $\left( P , Q_0 \right]$ cannot
affect the probability of producing data at $Q_1$ consistent with a
successful unveiling of either bit value $i$ there.  In particular, if
she follows the instructions of strategy $\St_0$ on $\left( P , Q_0
\right]$, and the instructions of strategy $\St_1$ on $\left( P , Q_1
\right]$, she has probabilities $p_i$ of producing data consistent
with a successful unveiling of bit value $i$ at $Q_i$, and hence
probability at least $\delta$ of producing data consistent with a
successful unveiling of bit value $0$ at $Q_0$ {\it and} with a
successful unveiling of bit value $1$ at $Q_1$.  

This means that, with
probability at least $\delta$, by combining her data at $Q_0$ and $Q_1$
at some point in their joint causal future, Alice can produce data
consistent with both sets of measurements in complementary bases.  
Thus, for example, for each state $\ket{\psi_i}$, she can identify
a subset of $2$ states from $\{ \ket{0}, \ket{1}, \ket{+}, \ket{-}
\}$, one from each basis, which must include $\ket{\psi_i}$.  

\lemma \qquad Given a single BB84 state $\ket{\psi}$, randomly chosen from
the uniform distribution, unknown to her, Alice's probability $p$
of choosing one of the subsets $S_1 = \{ \ket{0}, \ket{+} \}$, $S_2 = \{ \ket{+},
\ket{1} \}$, $S_3 = \{ \ket{1} , \ket{-} \}$, $S_4 \{ \ket{-} , \ket{0} \}$, 
that includes $\ket{\psi}$, is bounded by $p \leq \frac{1}{2} ( 1 +
\frac{1}{\sqrt{2}} )$, for any strategy.   An optimal 
strategy which realises this bound is to carry out the POVM 
\begin{equation}\label{optimalpovm}
\{ \frac{1}{2} P_1 , \frac{1}{2} P_2 , \frac{1}{2} P_3 , \frac{1}{2}
P_4 \}
\end{equation}
where $P_i$ is the projection onto the qubit $\ket{\phi_i} = 
\cos ( \theta_i ) \ket{0} + \sin ( \theta_i ) \ket{1} $,
$\theta_i = i ( \pi / 4 ) - ( \pi / 8 ) $, and 
given the outcome $P_i$, she guesses the subset $S_i$.  

\proof \qquad 

Recall first the standard state discrimination problem, in which Bob
chooses a state from the set $\{ \hat{\sigma}_j \}$ with associated
probabilities $\{ p_j \}$.  Alice later makes a measurement to try to
determine the state.  Her measurement may be described by a POVM $\{
\hat{\pi}_j \}$, where outcome $\hat{\pi}_j$ leads her to choose state
$\hat{\sigma}_j$ \footnote{Note that the number of outcomes does
  not have to equal the number of states prepared - in the general
  case there may not be a POVM element for every $j$.}.  The
probability that Alice identifies the state correctly is 
\begin{equation}
{\rm P_{corr}} = \sum_j p_j {\rm Tr}\left( \hat{\sigma}_j \hat{\pi}_j
\right) \, . 
\label{pcorr}
\end{equation}

In the variation here, Bob prepares a random state from the BB84 set.
Alice gets two guesses at the state.  These guesses must be
non-orthogonal BB84 states.  If
either guess is correct, she wins.  

Write the BB84 states as follows
\begin{equation}
\ket{e_1} = \ket{0} \, , 
\ket{e_2} = \ket{+} \, , 
\ket{e_3} = \ket{1} \, , 
\ket{e_4} = \ket{-} \, . 
\end{equation}
We use these states and the corresponding density matrices, $\hat{\rho}_j = \ket{e_j}
\bra{e_j}$, interchangeably below.  
Alice makes a measurement on the state received, and as a result of
the measurement makes her two guesses.  
Alice's possible guesses are 
$S_i = \{ \ket{e_i}, \ket{e_{i+1}} \}$ for $i=1$ to $4$.   Here and below 
we use the convention that $\ket{e_{i+4}} = \ket{e_{i}}$
and similarly for $\rho_i$ and $\hat{\pi}_i$.  
Each possible guess should
correspond to a measurement outcome, so we can associate each guess
$\{ \ket{e_i}, \ket{e_{i+1}} \}$ with a POVM element
$\hat{\pi}_{i}$ (some of which might in principle be zero).  
Now the probability
that Alice wins is the average over $i$ of the
probability that Alice chooses a set containing $\hat{\rho}_i$,
weighted by the probability that state $\hat{\rho}_i$ was prepared.
Explicitly we obtain:
\begin{eqnarray}
{\rm P(win)} &=& \frac{1}{4} \sum_i {\rm Tr}(\hat{\rho}_i
(\hat{\pi}_{i} +  \hat{\pi}_{i-1})) \\
&=& \frac{1}{2} {\rm Tr}( \frac{1}{2} ( \hat{\rho}_i +  \hat{\rho}_{i+1})
\hat{\pi}_{i}))  \, . \nonumber
\end{eqnarray}
Thus the problem is equivalent, up to a factor of $2$, to that of discriminating between the 
set of equiprobable states $\{ \frac{1}{2} ( \hat{\rho}_i
+ \hat{\rho}_{i+1} ) \}$.
Maximising the probability that Alice wins is equivalent to minimising the probability of error in discriminating these states.

A necessary and sufficient condition on a POVM realising a minimum
error measurement distinguishing general normalised states
$\hat{\sigma}_j$ chosen with probabilities $p_j$ is given by
\cite{Holevo73,Yuen75,Barnett09} 
\begin{equation}
\label{hel2}
\hat{\Gamma} - p_j\hat\sigma_j  \ge  0 \quad \forall j \, , 
\end{equation}
where
\begin{equation}
\hat{\Gamma} = \sum_i p_i\hat\sigma_i\hat\pi_i
\end{equation}
and ${\rm P_{corr}}= {\rm Tr}(\hat{\Gamma})$.  

For our transformed state discrimination problem we must calculate the operator
\begin{equation}
\hat{\Gamma} = \frac{1}{4} \sum_{i=1}^4 (\frac{1}{2} \hat{\rho}_i+
\frac{1}{2} \hat{\rho}_{i+1}) \hat{\pi}_{i} \, . 
\end{equation}
and show that
\begin{equation}
\hat{\Gamma} - \frac{1}{8} \hat{\rho}_{i} - \frac{1}{8} \hat{\rho}_{i+1}
\ge 0, \quad {\rm for~}i=1,2,3,4 \,. 
\end{equation}
It is straight-forward to verify that the POVM (\ref{optimalpovm}) satisfies this condition.
For this set
\begin{equation}
\hat{\Gamma} = \frac{1}{8}(1+\frac{1}{\sqrt{2}})\hat{I} \, . 
\end{equation}
Allowing for the factor of $2$ above, 
we obtain Alice's optimal guessing probability for the original problem as 
\begin{equation}
{\rm P_{win}} = 2 \Tr ( \hat{\Gamma} ) = \frac{1}{2}(1+\frac{1}{\sqrt{2}}) \, . \qquad {\rm QED} 
\end{equation}

\lemma \qquad Suppose now Alice is given a sequence of i.i.d. BB84 
states $\ket{\psi_i}_{i=1}^N$, randomly chosen from
the uniform distribution, and unknown to her, and is allowed to
perform a strategy $\St$ involving arbitrary collective operations.
Let $p_{i_1 , \ldots , i_{N-1} ; j_1 , \ldots , j_{N-1} }$
be her probability 
of choosing a subset from the list $S_1 = \{ \ket{0}, \ket{+} \}$, $S_2 = \{ \ket{+},
\ket{1} \}$, $S_3 = \{ \ket{1} , \ket{-} \}$, $S_4 = \{ \ket{-} , \ket{0} \}$, 
that includes the BB84 state $\ket{\psi_N}$, conditioned on the first $(N-1)$ states
supplied being $\ket{e_{i_1}}, \ldots , \ket{e_{i_{N-1}}}$ and her
guesses being $S_1 , \ldots , S_{j_{N-1}}$ respectively, where the
strategy $\St$ 
implies this is a possible list of guesses for the inputs. 
Then  $p_{i_1 , \ldots , i_{N-1} ; j_1 , \ldots , j_{N-1} } \leq \frac{1}{2} ( 1 +
\frac{1}{\sqrt{2}} )$, for any
  strategy $\St$ and any $\{ i_1 , \ldots , i_{N-1} ; j_1 \ldots
  j_{N-1} \}$
consistent with $\St$. 

\proof \qquad 

Suppose some collective strategy $\St$ violated this bound for some values
$\{ i_1 , \ldots , i_{N-1} ; j_1 , \ldots , j_{N-1} \}$.   Alice could then proceed as follows.
\begin{enumerate}
\item Prepare an entangled singlet state of two qubits, 
\item Prepare $(N-1)$ BB84 states $\ket{e_{i_1}}, \ldots ,
  \ket{e_{i_{N-1}}}$.  
\item Apply
strategy $\St$  (ignoring her knowledge of the BB84 states prepared) to the $(N-1)$ BB84 states and one qubit of the 
entangled states,
\item For the first $(N-1)$ states, check the guesses produced by $\St$,
\item If the results do not agree with $\{ S_{j_1} , \ldots ,
S_{j_{N-1}} \}$, return to step $1$ with a new singlet and a new batch
of BB84 states.   If they do agree, proceed to step $6$.
\item Apply a teleportation operation on the unknown BB84 state
$\ket{\psi_N}$ and the other singlet qubit, obtaining teleportation
unitary $U$.  Complete the
implementation of strategy $\St$, obtaining a guess at a subset
containing the teleported unknown qubit $U \ket{\psi_N}$. 
Apply the inverse $U^{\dagger}$ to obtain a guess at a subset $S_i$
containing $\ket{\psi_N}$.  By assumption, this guess is correct 
with probability $p_{i_1 , \ldots , i_{N-1} ; j_1 , \ldots , j_{N-1} } >  \frac{1}{2} ( 1 +
\frac{1}{\sqrt{2}} )$. 
\end{enumerate}

This iterated strategy is bound to proceed to step 6
eventually, and $\ket{\psi_N}$ is left isolated until step
6 is reached.  Alice thus has a strategy that produces
a subset guess for any single unknown state $\ket{\psi_N}$,
with success probability $p > \frac{1}{2} ( 1 +
\frac{1}{\sqrt{2}} )$,
contradicting Lemma 1.  QED

\theorem \qquad  Alice's probability $p_N$ of being able to produce data consistent
with measurements in complementary BB84 bases for $N$ random uniformly
i.i.d. unknown BB84 states obeys $p_N \leq (  \frac{1}{2} ( 1 +
\frac{1}{\sqrt{2}} ) )^N$.   
Hence, the security parameter $\delta$ in the bit commitment
protocol above obeys $\delta \leq ( \frac{1}{2} ( 1 +
\frac{1}{\sqrt{2}} ) )^N$.   

\proof follows from Lemma 2. 

Note that this
argument easily extends to give security bounds for large $N$ in the
presence of noise and errors, so long as the total noise and error rate is below $( \frac{1}{2} -
\frac{1}{2 \sqrt{2}} )$.   To see this let $Z_l = \sum_{k=1}^l j_k - l 
 \frac{1}{2} ( 1 +
\frac{1}{\sqrt{2}} )$, where $j_k = 1$ if Alice's subset guess on
the $k$-th state is correct and $j_k = 0$ otherwise.  We have $|Z_l| <
\infty$, $| Z_l - Z_{l-1} | \leq 
 \frac{1}{2} ( 1 + \frac{1}{\sqrt{2}} )
 $ 
and (from Lemma 2) $E(Z_l | W_k ) \leq Z_k$ for all $l>k$, where $W_k$ is the set of
subset guess outcomes up to state $k$.   So $Z_l$ is a supermartingale
and the Azuma-Hoeffding inequality implies 
\begin{equation}
{\rm Prob} ( \sum_{k=1}^N j_k \geq N ( 
\frac{1}{2} ( 1 +
\frac{1}{\sqrt{2}} ) + \epsilon )) \leq \exp (-N \epsilon^2 /( 2  (   \frac{1}{2} ( 1 +
\frac{1}{\sqrt{2}} ))^2 ) \, , 
\end{equation}
for any $\epsilon > 0$.  

{\bf Notes} 1. After the work reported above was completed, an independent security
analysis following different arguments was circulated by Kaniewski et al. \cite{kthw}. 

2. An alternative proof of Lemmas 1 and 2 and Theorem 1 follows by noting that for any
collective guessing strategy of Alice's, 
any particular subset guess $S_i$ for the $N$-th state, conditioned on
input
states $\ket{e_{i_1}}, \ldots , \ket{e_{i_{N-1}}}$ 
and guesses $\{ S_{j_1} , \ldots , S_{j_{N-1}} \}$ for the first $(N-1)$ guessing games,
must be represented by some positive operator $A = A^{\dagger} \geq 0$ 
on the $N$-th state.  Since the states are i.i.d. and uniformly 
distributed, the probability this guess is correct is

\begin{equation}
\Tr ( A ( \frac{1}{4} ( \hat{\rho_i} + \hat{ \rho_{i+1} } ))) / (
\frac{1}{2} \Tr (A) ) \, , 
\end{equation}
which is easily seen to be bounded by
$\frac{1}{2} ( 1 +
\frac{1}{\sqrt{2}} )$, for any value of $i$.  
That is, Alice's maximum confidence quantum
measurement \cite{cabgj} on the $N$-th state is unaltered if she
carries out 
collective measurements. 

Moreover, this implies a further security result.  Alice's
maximum confidence measurement on the $N$-th state cannot improve
on this success bound even if her strategy allows her sometimes to 
make no guess on some states (possibly including the $N$-th).  Hence the
protocol remains secure for large $N$ in the presence of any 
loss level (as reported by Alice) below $1$.       That is, it remains
secure even if Alice is allowed to report a large fraction of her
measurements as giving no result, so long as she tells Bob at
(essentially) the point $P$ which measurements were successful.   

3. Another proof of Theorem $1$ is given by verifying that a minimum
error measurement for $N$ BB84 states is obtained by taking the
$N$-fold tensor product of the POVM (\ref{optimalpovm}).  This  
follows straightforwardly using the method given above for $N=1$.  
\acknowledgments

AK was partly supported by a Leverhulme Research Fellowship and
a grant from the John Templeton Foundation.  SC and AK were partly
supported by 
Perimeter Institute for Theoretical Physics. Research at Perimeter
Institute is supported by the Government of Canada through Industry
Canada and by the Province of Ontario through the Ministry of Research
and Innovation.  AK thanks Serge Massar for helpful conversations.


\end{document}